\documentclass[%
 reprint,onecolumn,
 amsmath,amssymb,
 aps,
]{revtex4-2}
\usepackage{graphicx}
\usepackage{dcolumn}
\usepackage{bm}
\usepackage{import}
\usepackage{float}
\usepackage{caption}
\usepackage{subcaption}
\usepackage{booktabs}
\usepackage{tabularx}

\begin{document}

\preprint{APS/123-QED}

\title{Maximum Entropy Approach for the Prediction of Urban Mobility Patterns}

\author{Simone Daniotti}
 \altaffiliation[]{Complexity Science Hub Vienna, Vienna, 1080, Austria}
 \email{daniotti@csh.ac.at}
\author{Bernardo Monechi}%
\author{Enrico Ubaldi}%

\date{\today}

\begin{abstract}
The science of cities is a relatively new and interdisciplinary topic. It borrows techniques from agent-based modeling, stochastic processes, and partial differential equations. However, how the cities rise and fall, how they evolve, and the mechanisms responsible for these phenomena are still open questions. Scientists have only recently started to develop forecasting tools, despite their importance in urban planning, transportation planning, and epidemic spreading modeling. Here, we build a fully interpretable statistical model that, incorporating only the minimum number of constraints, can predict different phenomena arising in the city. Using data on the movements of car-sharing vehicles in different Italian cities, we infer a model using the Maximum Entropy (MaxEnt) principle. With it, we describe the activity in different city zones and apply it to activity forecasting and anomaly detection (e.g., strikes, and bad weather conditions). We compare our method with different models explicitly made for forecasting: SARIMA models and Deep Learning Models. We find that MaxEnt models are highly predictive, outperforming SARIMAs and having similar results as a Neural Network. These results show how relevant statistical inference can be in building a robust and general model describing urban systems phenomena. This article identifies the significant observables for processes happening in the city, with the perspective of a deeper understanding of the fundamental forces driving its dynamics.
\end{abstract}

\maketitle


\section*{Introduction}

In recent years, pressing societal and environmental problems such as population growth, migrations, and climate change, boosted the research on the Science of Cities and the related study of mobility.
The availability of large and detailed datasets covering the mobility of individuals at different granularity contributed largely to enhance the interest of researchers in this field\cite{bettencourt2007growth,bettencourt2010urban}.
Being multi-disciplinary, Science of Cities studies embrace diverse areas of research.
For example, statistical methods have been applied to city growth \cite{verbavatz2020growth}, multi-layer networks to urban resilience \cite{rogov2018urban} and spatial networks to describe and characterize the structure and the evolution of phenomena arising on it \cite{barthelemy2011spatial}.
On the modeling side, co-evolution models \cite{raimbault2020hierarchy} and agent-based simulations \cite{wise2016transportation} have been used to model stylized facts and take into account policy making.
Moreover, other frameworks such as ranking dynamics  ~\cite{noulas2012tale} have been used to study urban environments and universal laws arising in them.

The mobility patterns of individuals diffusing in an urban environment are determined by the interplay of different mechanisms, such as the daily routines of the individuals and environmental constraints, both again regulated by even more fundamental and interrelated phenomena, like wealth, trends, economic relations, socio-political disparities, and cultural movements \cite{davies_relatedness_2021,tainter1988collapse,schlapfer_universal_2021}. Urban environments are complex systems \cite{bettencourt2021introduction}, and describing their growth and relation with the surrounding cities is not unanimously understood by the scientific community \cite{verbavatz_growth_2020}.
Being that the commuting phenomenon inside and between cities may be driven by different events and may be related to different causes, different tools and understandings must be developed \cite{dong_universality_2022,noulas_tale_2012}.

In this work, we study urban mobility patterns, building a model based on only a few constraints driven by data observations. The model can predict different events in the urban environment with high precision and can be generalized to other dynamical processes unfolding in urban spaces.
Here, we propose a Statistical Inference (Maximum Entropy) approach to study and predict urban mobility patterns. 
Phenomena and relative observables that happen inside the city are complex, they entangle with various indicators and are difficult to predict.
To build a powerful, general, and robust statistical model, in principle we need to understand what are the important variables that play a central role in the phenomenon we want to model.
To solve it, we need to analyze the dataset we want to study, identify the important dynamical properties and then model it solving the problem of optimizing the resulting entropy.

The data represents the $30-minute-binned$ multi-variate time-series of the activity of different zones inside the city.

Being that urban systems are notoriously complex and that the fundamental causes of the observed mobility patterns are various and interrelated, our methodology is novel in this field in the sense that builds the most general model constrained to reproduce the correlations observed.

In section \textbf{Results} we present the results for the Metropolitan City of Milan: multi-variate forecasting of the zones activity and forecasting outliers events, discussed in \textbf{Discussion}.
We also settle the basis for the formal derivation of the model. We give an introduction to Maximum Entropy models, the formalism used in the following sessions and the optimization algorithm used.
Afterward,we introduce the formalism to compute an approximated Log-Likelihood. In \textbf{Additional Informations} can be found similar results for the cities of Rome, Florence, and Turin.

In section \textbf{Methods} we describe the data and the variables in use. To find out which are the important variables to represent, we do a historical analysis and observe a dynamics of contraction and dilation of the time-shifted correlations between different zones. With this, we understand the important effects and  and finally we define and derive the formulae for the ME model describing our data.

We use MaxEnt inference to obtain the parameters (using gradient ascent algorithm) and reproduce lag-correlations of definite positive time series.  We derive a model that is highly predictive at least as more sophisticated models that take into account nonlinear correlations and have more parameters. 
We also use the obtained statistical model to find extrema events (outliers, such as strikes and bad weather days).
As a result, we find the dynamics of cars' presence in urban areas to be extremely rich and complex. We infer the couplings parameters between the activity profiles of different areas and use them to project the cars' locations in time efficiently. 

\section*{Results}

\label{section:results}

In this section, given a brief introduction to the method and the data studied, we present the results.
For a more in depth description, we refer to sec. \textbf{Methods}.

\subsection*{Data}

\begin{figure}[H]
\centering
\includegraphics[width=17cm]{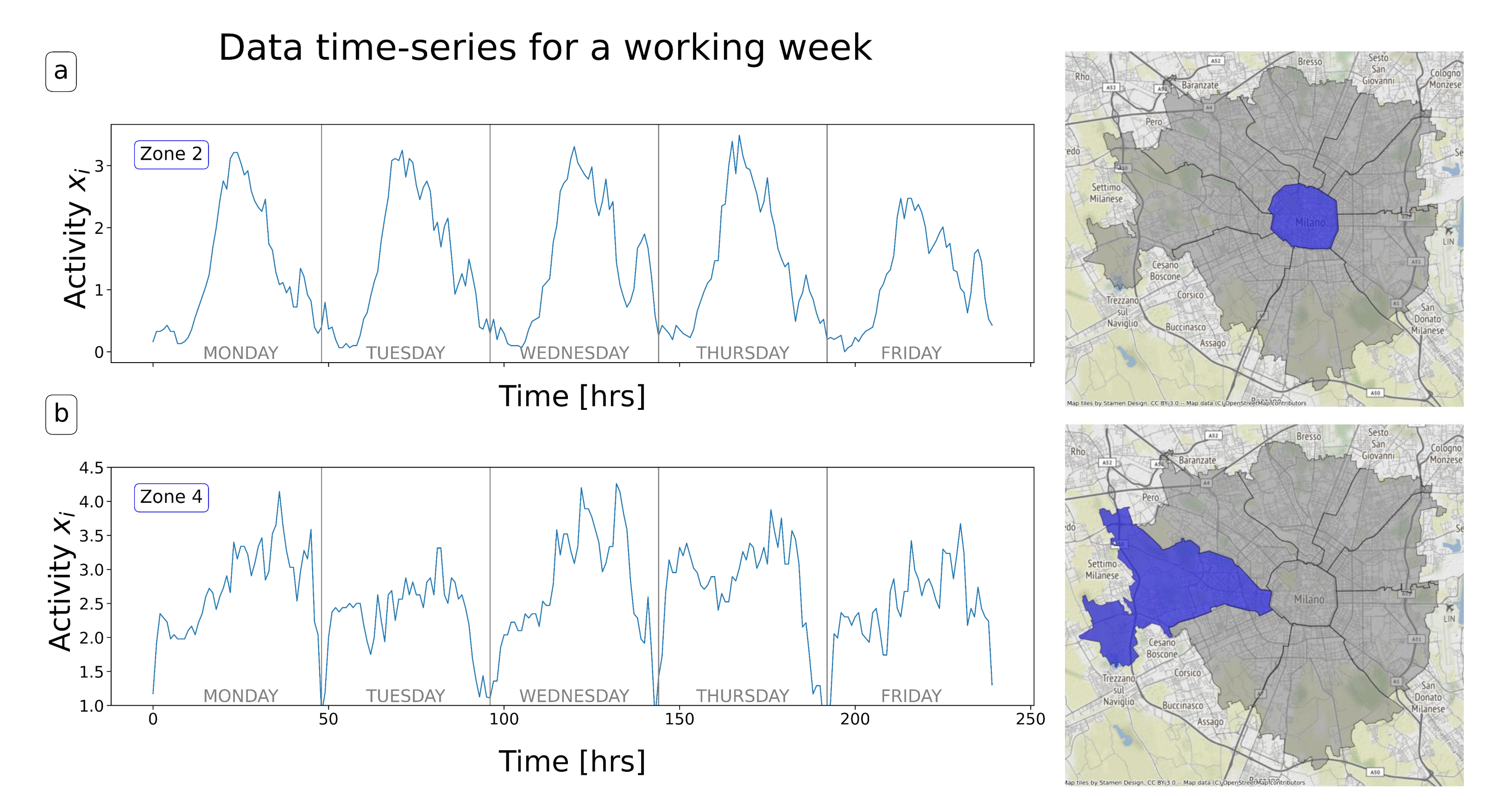}
\caption{Time series activity data for the city center (a) and for one of the suburbs (b).
\label{fig:19}}
\end{figure}

The data we use is a normalised multivariate time-series representing the parking activity of the different municipalities of the city.
The procedure to obtain the final form of the time-series in described in sec. \textbf{Methods}.
In Fig.~\ref{fig:19} we show an example of activity for two zones of our dataset for the Metropolitan City of Milan.

\subsection*{Maximum Entropy Principle}

\label{subsection:maxent}

The principle of maximum entropy states that the probability distribution which best represents the current state of knowledge is the one with the largest entropy, in the context of precisely stated prior data.
According to this principle, the distribution with maximal information entropy is the best choice. 
The principle was first shown by E. T. Jaynes in two papers in the late fifties where he emphasized a natural correspondence between statistical mechanics and information theory~\cite{jaynes1957information,jaynes1957information2}. 

In particular, Jaynes offered a new and very general rationale on why the Gibbsian method of statistical mechanics works. He showed that statistical mechanics, and in particular Ensemble Theory, can be seen just as a particular application of information theory, hence there is a strict correspondence between the entropy of statistical mechanics and the Shannon information entropy.

 Maximum Entropy models unveiled interesting results throughout the years for a large variety of systems, like flocking birds~\cite{bialek2012statistical}, proteins~\cite{ekeberg2013improved}, brain~\cite{tang2008maximum} and social systems~\cite{gresele2017maximum}.

 We will then implement this approach to define the model of our real-world system in the following sections. A more general introduction to the maximum entropy formalism is out of scope here and we refer to the existing literature for details~\cite{mehta2019high,giffin2009maximum,ibanez2019unsupervised,goodfellow2016deep}.

The probability distribution with Maximum Entropy $P_{ME}$ results from the extreme condition of the so-called \textit{Generalized Entropy}:

\begin{equation}
\label{eq:gen_ent}
\mathcal{S} \big[ P \big] = S \big[ P \big] + \sum_{k=1}^K \theta_k (\langle O_k \rangle_{P(\underline{X} )} -\langle O_k \rangle_{obs}),
\end{equation}
where 
\begin{equation}
S \big[ P \big]= - \sum_{\underline{X}} P(\underline{X}) \log(P(\underline{X}))
\end{equation}
 is the Shannon Entropy of the probability distribution $P(\underline{X})$.
The maximum of the Generalized Entropy is the maximum of the Entropy of the model when it is subject to constraints. Computing the functional-derivative (\ref{eq:gen_ent}) with respect to $P(\underline{X})$ and equating to zero results in:

\begin{equation}
\label{eq:pme}
P_{me}(\underline{X}) = \frac{1}{Z(\underline{\theta})} \exp \Big[- \sum_{k=1}^K \theta_k O_{k}(\underline{X})\Big],
\end{equation}
where
\begin{equation}
Z(\underline{\theta})=\int\limits_\Omega d\underline{X} \exp \Big[- \sum_{k=1}^K \theta_k O_{k}(\underline{X})\Big]
\end{equation}
is the normalization (making a parallel with statistical physics, can be called \textit{Partition Function}). $Z(\underline{\theta})$ is written as a sum if $\Omega$ is discrete.
Hence, the maximum entropy probability distribution is a Boltzmann distribution in the canonical ensemble at temperature $T = 1K$, with effective Hamiltonian $\mathcal{H}(\underline{X}) = -\sum_{k=1}^K \theta_k O_{k}(\underline{X})$.

It must be noticed that the minimization of the generalized entropy is equivalent to the maximization of the experimental average of the likelihood:

\begin{equation}
\mathcal{S} \big[ P \big] = \log Z(\underline{\theta}) - \sum_{k=1}^K \theta_k \langle O_k \rangle_{e} = -\langle \log P_{me} \rangle_{e} = \frac{1}{M} \sum_{m=1}^M \log P(\underline{X}^{(m)}).
\end{equation}

In other words, the $\theta_k$ are chosen by imposing the experimental constraints on Entropy or, equivalently, by maximizing the global, experimental likelihood according to a model with the constraints cited above.
Focusing on this last sentence, we can say that the optimal parameters of $\theta$ (called \textit{effective couplings}) can be obtained through Maximum Likelihood, but only once one has assumed (by the principle of Maximum Entropy) that the most probable distribution has the form of $P_{ME}$.

 Given the generative model probability distribution of configurations $P(\underline{X} \mid \underline{\theta})$ and its corresponding partition function by $\log Z( \underline{\theta} )$, the estimator of $\theta$ can be found by maximizing the log-likelihood:

\begin{equation}
\label{eq:loglike}
\mathcal{L}( \underline{\theta} ) = \langle \log(P(\underline{X} \mid \underline{\theta})) \rangle_{data} = - \langle \mathcal{H}(\underline{X} ; \underline{\theta}) \rangle_{data} - \log Z( \underline{\theta} ).
\end{equation}

Having chosen the log-likelihood as our cost function, we still need to specify a procedure to maximize it with respect to the parameters.

One common choice that is widely employed when training energy-based models is Gradient Descent~\cite{goodfellow2016deep} or its variations: optimization is taken w.r.t. the gradient direction.
Once one has chosen the appropriate cost function $\mathcal{L}$, the algorithm calculates the gradient of the cost function concerning the model parameters. The update equation is:
\begin{equation}
\theta_{ij} \leftarrow \theta_{ij} -\eta_{ij} \frac{\partial \mathcal{L}}{\partial \theta_{ij}}.
\end{equation}

Typically, the difficulty in these kind of problems is to evaluate $\log Z( \underline{\theta} )$ and its derivatives. The reason is that the partition function is rarely an exact integral, and it can be calculated exactly only in a few cases. However, it is still possible to find ways to approximate it and compute approximated gradients. 

\subsection*{Pseudo-Log-Likelihood (PLL) Maximization}
\label{subsection:pll}
Pseudo-likelihood is an alternative method compared to the likelihood function and leads to the exact inference of model parameters in the limit of an infinite number of samples \cite{arnold1991pseudolikelihood,nguyen2017inverse}. Let's consider the log-likelihood function $\mathcal{L}( \underline{\theta} ) = \langle \log P(\underline{X} \mid \underline{\theta}) \rangle_{data}$. In some cases we are not able to compute the partition function $Z( \underline{\theta} )$, but it is possible to derive exactly the conditional probability of one component of $\underline{X}$ with respect to the others, i.e. $P(X_j | \underline{X_{-j}},  \underline{\theta})$ where $\underline{X_{-j}}$ indicates the vector $\underline{X}$ without the $j$-th component.

In this case, we can write an approximated likelihood called Pseudo-log-likelihood, which takes the form:
\begin{equation}
\label{eq:pseudo}
\mathcal{L}( \underline{\theta} )_{\textit{pseudo}} = \sum_j \langle \log P(X_j | \underline{X_{-j}}, \underline{\theta}) \rangle_{data}.
\end{equation}

The model we will introduce in this work does not have an explicit form for the partition function, but Eq.~(\ref{eq:pseudo}) and its derivatives can be calculated exactly. Thus, the Pseudo-log-likelihood is a convenient cost function for our problem.

\subsection*{Definition of MaxEnt model}
\label{subsection:definition}
We have seen that the Maximum Entropy inference scheme requires the definition of some observables that are supposed to be relevant to the system under study. Being that one aims to predict the evolution of the different $x_i(t)$, the most simple choice is to consider their correlations.

As a preliminary analysis, we study the correlation between the activity $x_i(t)$ of most central zone within the city (highlighted in red) and all the other zones with a \textit{shift} in time (i.e. we correlate $x_i(t)$ with $x_j(t-\delta)$ for al $j\neq i$ and some $\delta>0$). The measure of correlation between two vectors $u$ and $v$ represented is $\frac{u \cdot v}{{||u||}_2 {||v||}_2}$, where ${||.||}_2$ measures the euclidean norm defined in $\mathbb{R}^n$ as $\left\| \boldsymbol{x} \right\|_2 := \sqrt{x_1^2 + \cdots + x_n^2}$.
We see that the areas with large correlations vary with $\delta$ so that they clustered around the central area for small values, and they become peripherals when $\delta \sim 31$. When $\delta \sim 48$ they start to cluster around the central area again. 
An opposite behavior is shown when repeating the same procedure focusing on a peripheral zone.
We perform a historical analysis of time-shifted correlations, observing contraction and dilation w.r.t. different zone in a one-day periodicity.
Due to the correlation dependency on the time shift, we have to include it in the observables used to describe the system. Hence, we chose as observables all the \textit{shifted correlations} between all the couples of zones defined as,
\begin{equation}
\label{eq:corr}
\langle x_i(t) x_j(t-\delta) \rangle_{data} = \frac{1}{T-\Delta} \sum_{t=\Delta}^T x_i(t)  x_j(t-\delta),
\end{equation}
for $i,j=1,....,N$ (with $N$ as the number of zones we took in consideration) and $\delta=0,....,\Delta$.
Another common choice is to also fix the average value of all the variables of the system. We took $\langle x_i \rangle_{data} = \frac{1}{T-\Delta} \sum_{t=\Delta}^T x_i(t)$ and 
$\langle x^2_i \rangle_{data} = \frac{1}{T-\Delta} \sum_{t=\Delta}^T x^2_i(t)$.

From these, we obtain the equation for the probability:

\begin{equation}
P(x(t), \dots, x(t-\Delta)) = \frac{1}{Z} \exp \left[ -\sum_{t=\Delta}^T \sum_i a_i x_i^2(t) + \sum_{t=\Delta}^T \sum_i h_i x_i(t) + \sum_{t=\Delta}^T \sum_{\delta=1}^\Delta J_{ij}^\delta x_i(t)x_j(t-\delta) \right ],
\end{equation}

where $a_i$ is the $i$-th component's standard deviation, $h_i$ is its mean and $J_{ij}^\delta$ are the time shifted interactions.

Writing $v_i(t) = h_i  + \sum_\delta J_{ij}^\delta x_j(t-\delta)$ one can obtain:

\begin{equation}
P(x(t), \dots, x(t-\Delta)) = \frac{1}{Z} \prod_{t=\Delta}^T \exp \left[ -\sum_i a_i x_i^2(t) + \sum_i v_i(t) x_i(t) \right ].
\end{equation}

\subsection*{Time series Forecasting}

We trained the model for $100000$ steps using several hyperparamenters ($\Delta = \{24,36,48,72\}$ and $\lambda=\{0.001,0.004,0.005,0.006,0.01\}$, see tab.\ref{tab5}). 

To quantify the predictive capabilities of our models, we used the \textit{Mean Absolute Error (MAE)}, the \textit{Mean Squared Error (MSE)} and the \textit{Coefficient of Determination}, also known as $R^2$ \cite{devore2011probability}.

\begin{table}[H] 
\caption{$R^2$ values w.r.t. $\Delta$ and $\lambda$ values, resulting from $100000$ steps of training. \label{tab5}}
\newcolumntype{C}{>{\centering\arraybackslash}X}
\begin{tabularx}{\textwidth}{CCC}
\toprule
\textbf{$\Delta$}	& \textbf{$\lambda$}	& \textbf{$R^2$}\\
\midrule
24 & 0.001 & 0.855316 \\ 
24 & 0.004 & 0.854723 \\ 
24 & 0.005 & 0.860069 \\ 
24 & 0.006 & 0.853689 \\ 
24 & 0.01 & 0.832542 \\ 
36 & 0.001 & 0.854458 \\ 
36 & 0.004 & 0.85442 \\ 
36 & 0.005 & 0.853946 \\ 
36 & 0.006 & 0.853447 \\ 
36 & 0.01 & 0.832672 \\ 
48 & 0.001 & 0.860072 \\ 
48 & 0.004 & 0.861741 \\ 
48 & 0.005 & 0.869466 \\ 
48 & 0.006 & 0.861626 \\ 
48 & 0.01 & 0.833099 \\ 
72 & 0.001 & 0.85655 \\ 
72 & 0.004 & 0.859633 \\ 
72 & 0.005 & 0.867582 \\ 
72 & 0.006 & 0.859632 \\ 
72 & 0.01 & 0.858497 \\ 
\bottomrule

\end{tabularx}
\end{table}

\begin{figure}[H]
\centering
\includegraphics[width=13.5cm]{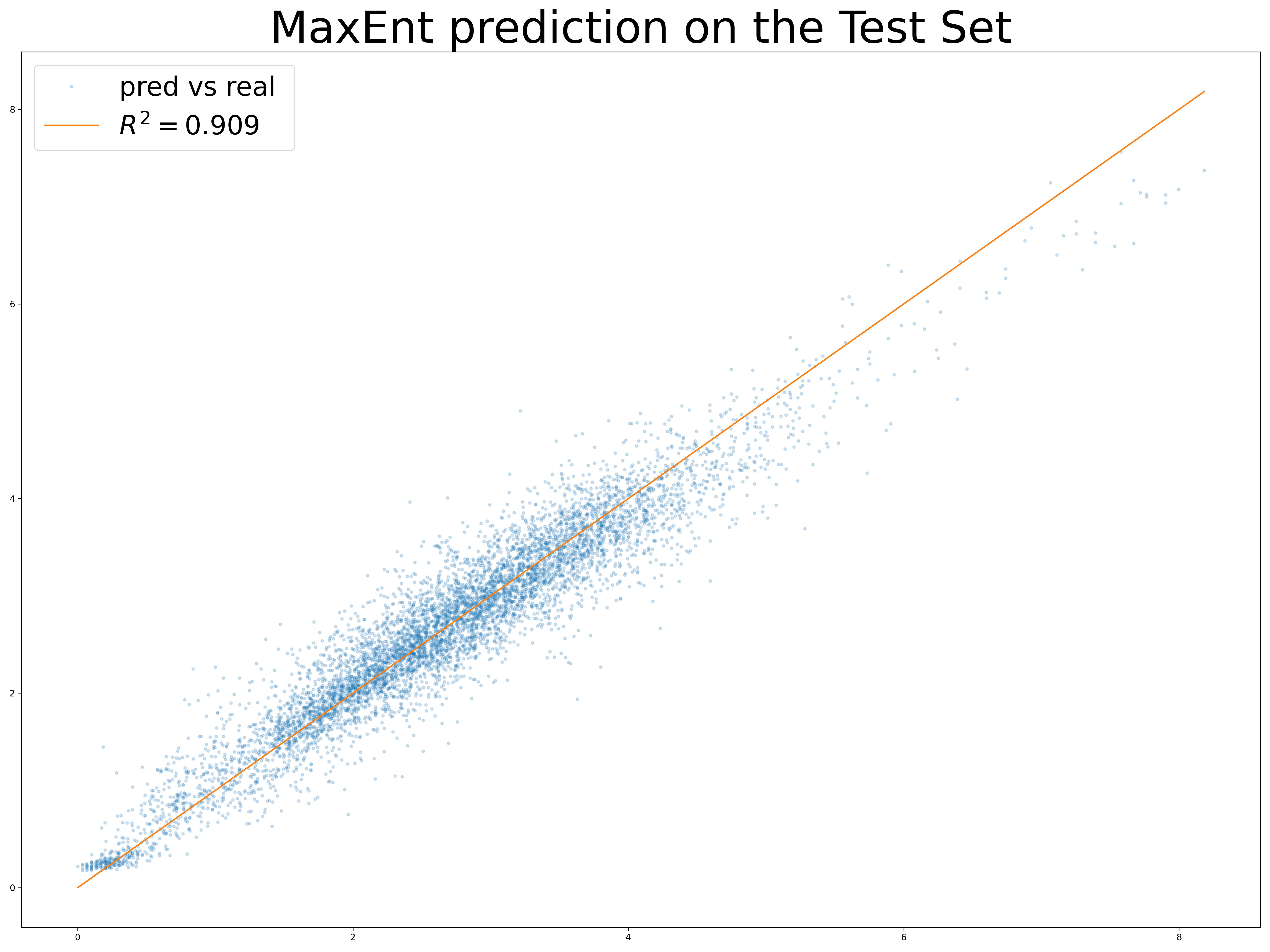}
\caption{R2 MaxEnt prediction over the test set. To have r2=1, the point should align on the bisector line, having a perfect predictor.
\label{r2}}
\end{figure}

We typically find good predictive capabilities on the test set ( as in fig.~\ref{r2} and the exact values for each predictor are found in fig~\ref{tab5}). To check for the best value, we used the average value at each time step during the day in the training set as a baseline model.  For $\Delta = 48$ (exactly the day-periodicity) and $\lambda=0.005$, we find our model's best performance, so we will use these values.

To assess the predictive power of our approach against more complex models, we compared it with standard statistical inference and Machine Learning methods. In particular:

\begin{enumerate}

\item a \textbf{SARIMA} model \cite{tseng2002fuzzy}: we performed a grid search over the parameters using \textit{AIC} \cite{burnham2004multimodel} as metrics, the best predictive model is, referring to standard litherature on the topic, $\{p=2,d=0,q=1,P=2,D=1,Q=2,m=48\}$. In tab.\ref{sarima_table} the table of the grid search. 
\item \textbf{NN} and \textbf{LSTM}: each of them has 3 hidden layers with a $N_{nodes}=\{64,128,256,512\}$. Between the layers, we performed dropout techniques and standard non-linear activation functions (ReLU).
Performing a grid-search over the other Machine Learning hyperparameters, the layers with $128$ and $256$ nodes perform better.

\end{enumerate}

\begin{figure}[H]
\centering
\includegraphics[width=13.5cm]{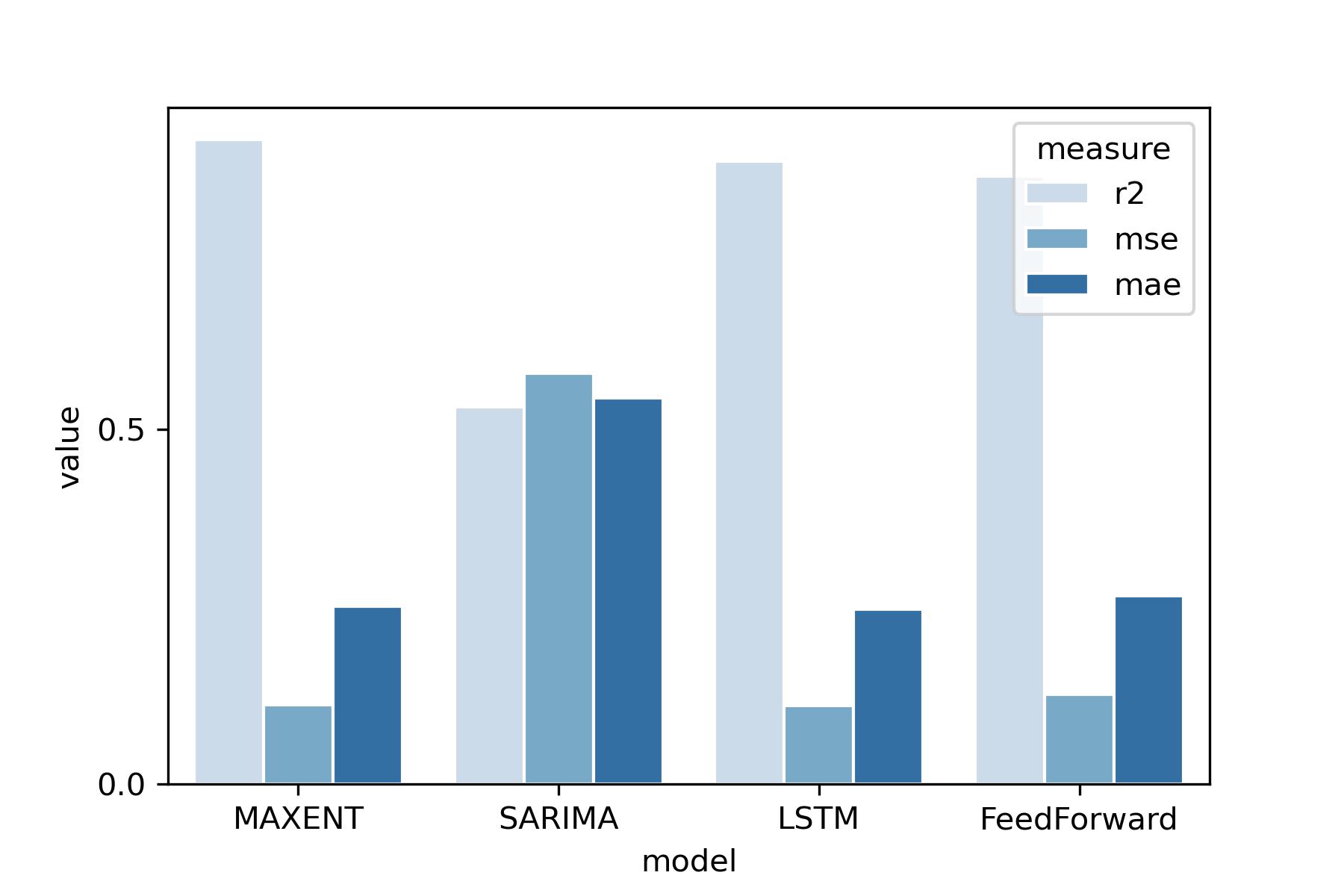}
\caption{ Model comparison using different metrics. We see similar results between the MaxEnt model and the Deep Learning ones.
\label{model_eval}}
\end{figure}

\begin{figure}[H]
\centering
\includegraphics[width=13.5cm]{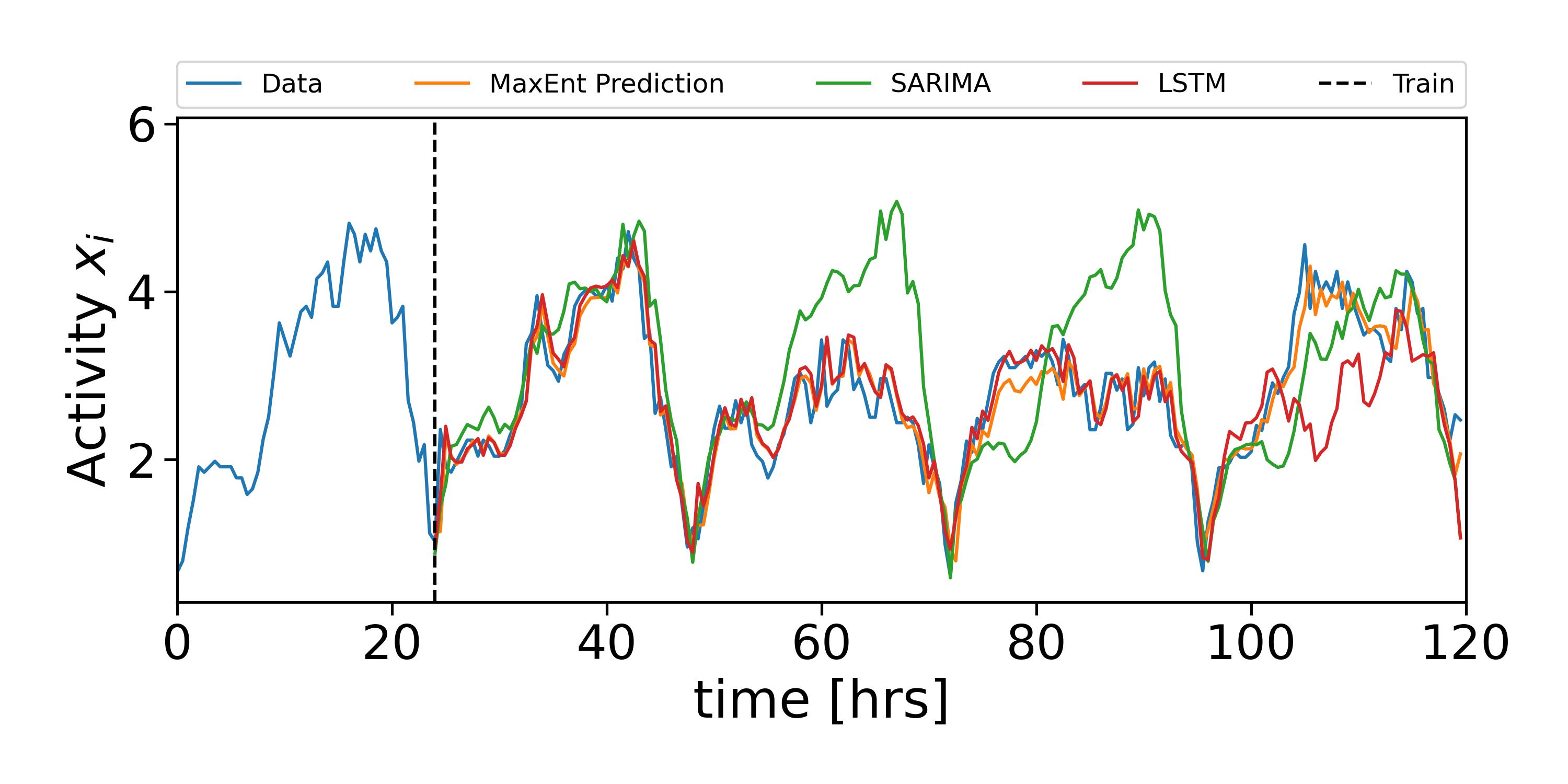}
\caption{Model comparison: taking one zone's time series, we see that SARIMA models are good predictors of periodicity, but the other models perform better when the change between one period and another is high.
\label{ts_pred}}
\end{figure}

Fig.~\ref{model_eval} shows that our model outperforms SARIMA and performs as well as the Machine Learning ones. Being that the LSTM one always performs slightly better than the FeedForward, we will use the first as a reference.  Fig.~\ref{ts_pred} shows this in more detail: SARIMA fails to reproduce variations and it is mostly focused on seasonality. Our model works as well as NNs but with two orders of magnitude of fewer parameters ($109$ w.r.t., for the $128$-nodes configuration, $39040$ for the FeedForward).

In fig. \ref{fig:cities} we show data and results for the cities of Rome,  Florence, and Turin.

\subsection*{Extreme events prediction}
\label{subsection:extreme}
\begin{figure}[H]
\centering
\includegraphics[width=13.5cm]{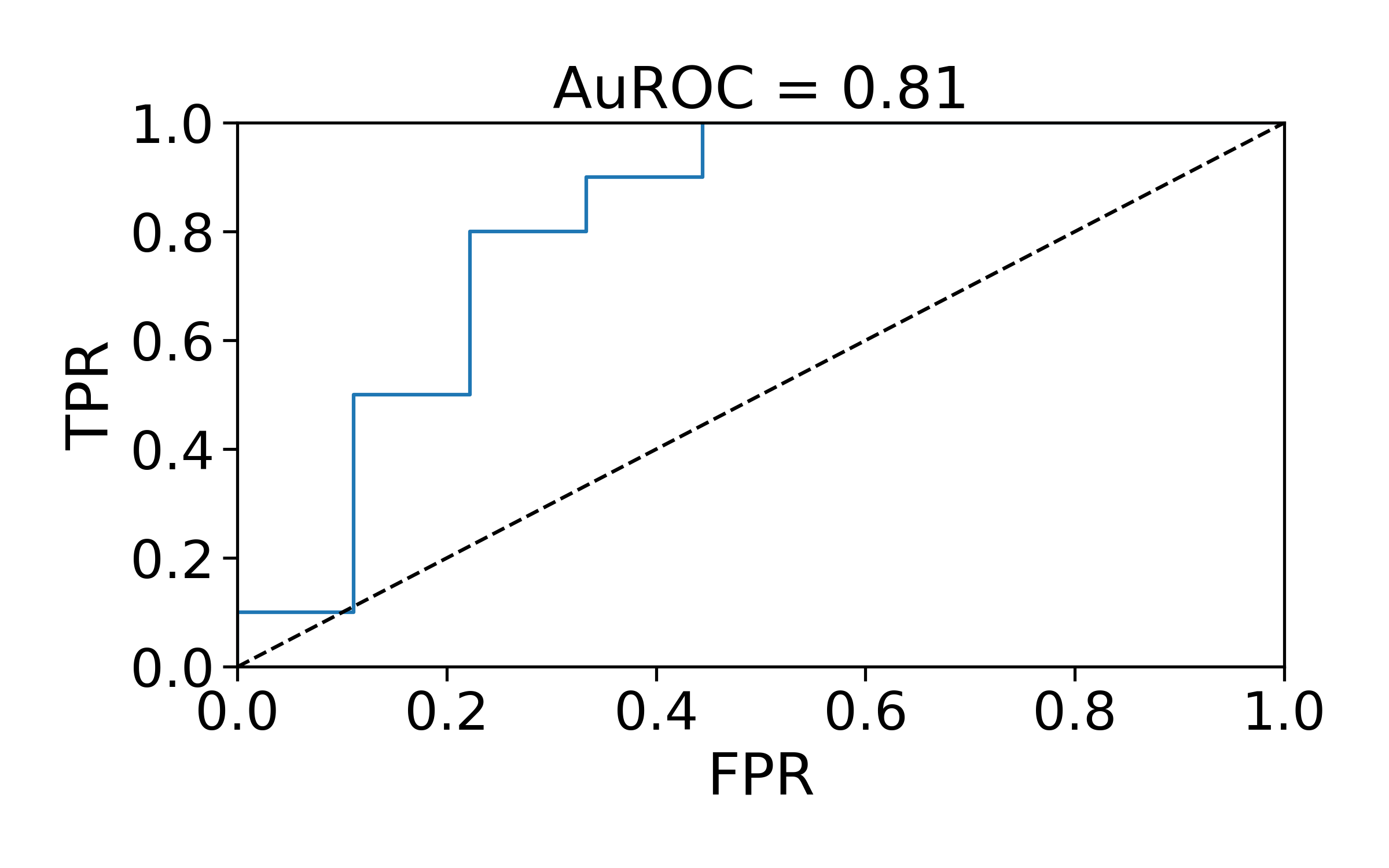}
\caption{ROC curve for the detection of outliers (bad weather conditions and strikes) for $\lambda=0.006$ and $\Delta=48$.\label{roc2}}
\end{figure}

As already pointed out, prediction is a possible application of our procedure, and we compared the results with state-of-the-art forecasting techniques. Another possible application is the detection of outliers in the data. We see from Fig.~\ref{fig:19} that our car sharing data exhibits quite regular patterns on different days. These patterns can be perturbed by a wide variety of external events, influencing the demand for car-sharing vehicles as well as the traffic patterns in the city. If we compute the log-likelihood from equation~(\ref{eq:pll}) restricting to one of the perturbed days, we would expect it to be particularly low, implying that the model has lower predictive power than usual. Hence, we can use the log-likelihood $\mathcal{L}$ from equation~(\ref{eq:pll}) as a score in a binary classification task. After training the model with a specific parameter choice, we consider a set of days in the test dataset, $d_1 \dots d_D$. Each one is a set of consecutive time steps: $d_i = [t^{d_i}_1, t^{d_i}_2]$ with $t^{d_i}_1 < t^{d_i}_2$. The log-likelihood of a single day is just
\begin{equation}
\label{eq:pll_singleday}
\mathcal{L}_{pseudo}(d_i) = \frac{1}{t^{d_i}_2 - t^{d_i}_1-\Delta} \sum_{t= t^{d_i}_1-\Delta}^{t^{d_i}_2} \log P(x(t) | x(t-1), \dots, x(t-\Delta)).
\end{equation}
We can now assign a label $L(d_i)=0$ if $d_i$ is a standard day and  $L(d_i)=1$ if it is a day where a certain external event occurred. We considered two types of external events:
\begin{enumerate}
    \item Weather conditions: on that day there was fog, a storm or it was raining.
    \item Strikes: in that day there was a strike in the city capable of affecting car traffic. We considered Taxi strikes, Public Transportation strikes, and general strikes involving all working categories. 
\end{enumerate}
We had $50\%$ of days labeled as 1 among the days in the test dataset. In fig.~\ref{roc2} we show the ROC curve for the classification of such days, using the log-likelihood trained with the parameters $\{\Delta=48,\lambda=0.006\}$. The Area under the ROC curve (AuROC) is 0.81, indicating that $\mathcal{L}$ is a good indicator of whether an external event occurred on a specific day.

\section*{Discussion}

In this work, we addressed the problem of building statistical and machine learning
models to predict mobility patterns within the Milan Metropolitan Area. We focused on the
prediction of how many car-sharing vehicles are parked within a specific area at different
times. To do this, we analyzed a longitudinal dataset of parking events coming from
the APIs of the Italian car-sharing company “Enjoy”. The processed data consisted of a
collection of time series representing the number of parked vehicles in a given area at a
given time.

To predict the evolution of these time series, we used a Maximum Entropy (ME)
approach that requires the identification of relevant observables from the data and use
these observables to define probability density functions depending on some parameters.

Maximum Entropy modeling has proven to be very effective at studying and determining the activity patterns inside the city (Fig.~\ref{ts_pred}).

We compared our model with other models specifically built for time series forecasting. In particular, we used a SARIMA model, a Feed-Forward Neural Network, and a Long Short-Term Memory Neural Network.

Maximum Entropy models outperformed SARIMA models w.r.t. all the metrics used in the evaluation.
Our model is as predictive as LSTMs Neural Networks, but using two orders of magnitude fewer parameters , interpretability of the parameters, and the possibility of performing different studies using the same model.

Finally, we used the statistical model also to identify extreme events affecting urban traffic, such as bad weather conditions or strikes.

In conclusion, this work is a first attempt at modeling the inherently complex interactions and causalities that shape the fluxes of people and vehicles within urban spaces. We showed that the interplay of the activity of different areas at a different time is far from trivial. Indeed, we found the models derived within the ME framework to accurately reproduce the empirical time series using a mixture of correlations between different areas at different times. 

Moreover, our finding that linear correlations are enough to predict the mobility patterns (wrt neural networks) is an interesting result that requires additional investigations.

In \textbf{Additional Information} we show the prediction ability on other three cities: Rome, Florence, and Turin.

Given our results, several research directions can be taken to improve and extend the results:
\begin{itemize}
\item  a more extensive study on the effects of seasonality could help in building better models. Specific season-dependent models could be built by taking into account larger datasets.
\item Evaluate how the prediction ability of the Maximum Entropy models is dependent upon the structure of the city or the resolution of the time series.
\item Entangling mobility patterns with other city indicators, such as socio-political disparities and economic indicators, can lead to a better model that depicts human distribution. 
\item The inclusion of non-linear interaction in the ME models could be hard if made by defining standard observables. Instead, \textit{hidden variables} approaches could be take into account, e.g. Restricted Boltzmann Machines \cite{sutskever2009recurrent}. 
\item The ME models could be adapted to perform other anomaly detection tasks, i.e. identifying parts of the time series which are not standard fluctuations of the system \cite{fiore2013network,sun2004mobility}.  
\end{itemize}

\section*{Methods}

\subsection*{Dataset}
\label{subsection:dataset}

The dataset used in this work is a log of the cars' locations, belonging to \textit{Enjoy}, a popular Italian car-sharing service. Although the service is active in six major Italian cities, we focus our analysis on the city of Milan. The data has been collected via the Enjoy APIs, scraping the information through a client account. We obtained information about the area of service (i.e., the limit where people can start and end their ride), the location of vehicles, and the points of interest (POIs) related to the service (such as fuel stations and reserved parkings) from the API endpoints. The endpoints used have base URL \textit{https://enjoy.eni.com/ajax/} and the following endpoints: \href{https://enjoy.eni.com/ajax/retrieve_area}{\href{retrieve\_areas}} for the area, \href{https://enjoy.eni.com/ajax/retrieve_vehicles}{\href{retrieve\_vehicles}} for cars, and \href{https://enjoy.eni.com/ajax/retrieve_pois}{\href{retrieve\_pois}} for points of interests.

We collected the data recording all the movements of Enjoy cars within the city of Milan in $2017$.  Through the scraping procedure, we collected a series of events, that are divided into two categories: parking events and travel events of each Enjoy car we could observe. Each event comes with a set of information such as the car plate, the time at which the event occurred (with the precision of seconds), and the latitude and longitude of the parking spot in the case of parking events. Starting and arrival points in the case of travel events are also recorded. In the rest of this article, we will focus mainly on parking events, but our methods and analyses could be also applied to study the volume of travel events.
The final aim of the work was to predict the number of vehicles parked at a given time in each and every district of the city. To this end, we divided the area of service into the different municipalities (we will call them \textit{zones}). Different tessellation strategies (e.g., using hexagons with a $900 m$ side) have been considered, but the municipality division provides more statistichally stable training data, and thus more precise models.

For each type of event (parked car)  that we collected, we defined the \textit{activity} of a zone as the number of events (i.e., cars parked) occurring there in a certain fixed amount of time $\delta t$. Indicating each zone with the index $i$, we have obtained for each and every one of them a discrete time series $x_i(t)$ representing the \textit{activity} of the zone $i$ in the time frame $[t, t+\delta t]$. We will indicate with $t=T$ the last time frame observed in the data. In the following, we will use a $\delta t=1800\,$s, corresponding to $30$ minutes.
This time bin width has been chosen so to have a stable average activity $\langle x_i(t)\rangle$ and a stable variance $\sigma(x_i(t))$  over all the zones. Indeed, narrower or larger time bins feature unstable mean and/or variance in time: the $30-minutes$ binning is significantly stable throughout all the observation period. This characteristic helps the model to generalize and predict in a better way as the distributions of the modeled quantities are not changing too much in time.

Some of the models we would like to implement require real variables distributed over $\mathbb{R}$. However, the $x_i(t)$ activity we have defined so far belongs to $\mathbb{N}$ by definition.
For this reason, we defined another kind of variable dividing the activity by the standard deviation:
\begin{equation}
z = \frac{x}{\sigma},
    \label{eq:zscore}
\end{equation}
where $x$ is the original activity data, and $\sigma$ is the standard deviation of the activity in time. In this case, for the population we take into account the \textit{typical day}: the std is taken w.r.t. the same time bin for every day.
So, Eq.~(\ref{eq:zscore}) gets into:
\begin{equation}
z_i(t)= \frac{x_i(t)}{\sigma_{i}(t_{\delta t})},
\end{equation}
where $t_{\delta t} = t\mod(\delta t)$ is the integer division of t by the $\delta t$ time bin width, $\mu_i(t_{\delta t})$ is the average over all the activities of the zone $i$ at times with the same value of $t \mod \delta t$, and $\sigma_{i}(t_{\delta t})$ is their standard deviation.
To keep the notation formalism, we will indicate $z_i(t)$ with $x_i(t)$, keeping in mind that now $x$ refers to a \textit{normalized activity}. 

From now on, we will work on the normalized $x_i(t)$: these indicate how much a given area is more "active" - i.e., has more cars parked - concerning the average volume of that time bin $t$ (typical day activity) by weighting this fluctuation with the standard deviation observed for that zone-hour volume. In this way, we can compare the signal of areas with high fluctuations and high activities with less frequented areas around the city.

The final step when processing data for inference and Machine Learning is to divide the data into a \textit{train set} and a \textit{test set}. The models will be then trained using the first dataset, and their precision will be tested on the second one so to check their ability to generalize to unseen data. Different kinds of splittings have been done, like random splitting or taking the first part of the time series as training and the last one as a test.
Similar results are obtained.

As a final remark, in the following we will indicate with $t$ the time bin of activity, i.e. $x_i(t)$ will indicate the activity of the zone $i$ in a time range $[t \delta t, (t+1)\delta t]$.

\subsection*{Pseudo-Log-Likelihood Maximization}
\label{sec:2_pll}

The Boltzmann probability is defined over the whole time series of all the zones, i.e. 

\begin{equation}
P \Big( x(t), x(t-1),...,x(t-\Delta) \Big) \propto \exp \Big( \mathbf{H}(x(t), x(t-1),...,x(t-\Delta)) \Big).
\end{equation}
From this, it is straightforward to define the conditional probability of one-time step $x(t)$ concerning all the previous ones:
\begin{equation}
\label{eq:conditioned}
P(x(t) \mid x(t-1),...,x(t-\Delta)) = \frac{P \Big( x(t), x(t-1),...,x(t-\Delta) \Big)}{P \Big( x(t-1),...,x(t-\Delta) \Big)}.
\end{equation}
Using equation (\ref{eq:pseudo}), we can define the Pseudo-Log-Likelihood as:
\begin{equation}
\label{eq:pll}
\mathcal{L}_{pseudo} = \frac{1}{T-\Delta} \sum_{t=\Delta}^T \log P(x(t) | x(t-1), \dots, x(t-\Delta)).
\end{equation}
Here, using Eq. (\ref{eq:conditioned}) and substituting the functional form of the two total probabilities, we obtain
\begin{equation}
\label{eq:pll_cond}
P(x(t) | x(t-1), \dots, x(t-\Delta)) =  \prod_i \frac{1}{Z_i(t)} \exp( -a_i x_i^2(t) + x_i(t) v_i(t)) ,
\end{equation}
with
\begin{equation}
\begin{aligned}
&Z_i(t) = \frac{1}{\sqrt{a_i}} \exp \left( \frac{v_i(t)^2}{4a_i} \right ) \int_{-\infty}^{v_i(t)/2 \sqrt{a_i}} dz e^{-z^2} = \frac{1}{\sqrt{a_i}} \exp \left( \frac{v_i(t)^2}{4a_i} \right ) I\left(\frac{v_i(t)}{2 \sqrt{a_i}} \right) , \\
&v_i(t) = \sum_{n=1}^{d} \sum_\delta  (J_n ^\delta x^n(t-\delta))_i + h_i.
\end{aligned}
\end{equation}

Substituting in eq. (\ref{eq:pll}), we get:
\begin{equation}
    \mathcal{L}_{pseudo} = \frac{1}{T-\Delta} \sum_{t=\Delta}^T \sum_i  -a_i x_i^2(t) + x_i(t) v_i(t) - \log Z_i(t)
\end{equation}
and we can calculate the gradients of the $\mathcal{L}_{pseudo}$ w.r.t. the parameters:
\begin{equation}
    \begin{aligned}
       &  \frac{\partial \mathcal{L}_{pseudo}}{\partial a_{i}} = \frac{1}{T-\Delta} \sum_t -x_i^2(t)  - \frac{\partial \log Z_i(t)}{\partial a_i},\\
       & \frac{\partial \mathcal{L}_{pseudo}}{\partial J^\delta_{ij}} = \frac{1}{T-\Delta} \sum_t x_i(t) x_j(t-\delta) -  \langle x_i(t) \rangle x_j(t-\delta),\\
       & \frac{\partial \mathcal{L}_{pseudo}}{\partial h_{i}} = \frac{1}{T-\Delta} \sum_t x_i(t)  -  \langle x_i(t) \rangle ,
    \end{aligned}
\end{equation}

where 
\begin{equation}
\langle x_i(t) \rangle = \frac{v_i(t)}{2a_i} + \frac{1}{2 \sqrt{a_i} I(\frac{v_i}{2 \sqrt{a_i}} )} \exp \left ( -\frac{v_i(t)^2} {4a_i}\right)
\end{equation}

and 

\begin{equation}\frac{\partial \log Z_i(t)}{\partial a_i} = -\frac{1}{2a_i} + \frac{-v_i^2(t)}{4a_i^2} + \frac{1}{I(\frac{v_i}{2 \sqrt{a_i}} )} \exp \left ( -\frac{v_i(t)^2} {4a_i}\right) \left ( \frac{-v_i(t)}{4 a_i^{3/2}} \right).
\end{equation}

The fact that the gradients and the cost function can be computed exactly makes the inference of the parameters relatively easy with respect to other cases where they need to be approximated.

Once the parameters of the model have been inferred with some method, it is possible to use it to predict the temporal evolution of the normalized activities of the system. Given some certain state of the system unit time $t-1$, i.e. $(x(t-1), \dots x(t-\Delta))$ (past time steps further than $\Delta$ from $t$ are not relevant), we can use equation (\ref{eq:pll_cond}) to predict the next step $x(t)$. Since the probability in (\ref{eq:pll_cond}) is a normal distribution whose average is completely defined by $(x(t-1), \dots x(t-\Delta))$, the best prediction of $x_i(t)$ is the average of the distribution $\langle x_i(t) \rangle = \frac{1}{2a_i} \left (  v_i(t) + \frac{1}{Z_i(t)} \right )$.

In other words, we are using the generative model to make a discrimination task. This makes possible the comparison of this model with standard machine learning ones, by comparing their precision in the prediction of the time series. To avoid over-fitting, we used L1-regularization~\cite{buhlmann2011statistics,bishop2006pattern}.
An in-depth description of the technique can be found in the references.
In practice, the cost function that has to be optimized is:

\begin{equation}
C(\theta) = \log P(\theta | X) = \log P( X | \theta ) - \lambda \sum_i |\theta_i| + cost .
\end{equation}

The first term of this sum is the Log-likelihood; instead, the second term is the regularization term.

If the gradient is performed, we obtain:

\begin{equation}
\frac{\partial C(\theta)}{\partial \theta_i} = \frac{\partial \log P( X | \theta )}{\partial \theta_i}  - \lambda sign(\theta_i),
\end{equation}

where $sign$ is the sign function.

The training curves show no sign of overfitting,as the log-likelihood asymptotically stabilizes for the validation and train sets (see Fig.~\ref{training} in Appendix).

\appendix

\section{Appendixes}

\begin{table}
\caption{Grid search over the SARIMA parameters\label{sarima_table}}
\newcolumntype{C}{>{\centering\arraybackslash}X}
\begin{tabularx}{\textwidth}{CCCCCCCC}
\toprule
\textbf{p}	& \textbf{d}    &   \textbf{q}  &   \textbf{P} & \textbf{D} & \textbf{Q}  & \textbf{m} & \textbf{AIC}\\
\midrule 
1 & 0 & 1 & 0 & 1 & 1 & 48 & inf \\
0 & 0 & 0 & 0 & 1 & 0 & 48 & 1243.245 \\
1 & 0 & 0 & 1 & 1 & 0 & 48 & 359.487 \\
0 & 0 & 1 & 0 & 1 & 1 & 48 & inf \\
0 & 0 & 0 & 0 & 1 & 0 & 48 & 1241.606 \\
1 & 0 & 0 & 0 & 1 & 0 & 48 & 615.697 \\
1 & 0 & 0 & 2 & 1 & 0 & 48 & 285.295 \\
1 & 0 & 0 & 2 & 1 & 1 & 48 & 270.777 \\
1 & 0 & 0 & 1 & 1 & 1 & 48 & 279.520 \\
1 & 0 & 0 & 2 & 1 & 2 & 48 & 237.403 \\
1 & 0 & 0 & 1 & 1 & 2 & 48 & inf \\
0 & 0 & 0 & 2 & 1 & 2 & 48 & inf \\
2 & 0 & 0 & 2 & 1 & 2 & 48 & 236.616 \\
2 & 0 & 0 & 1 & 1 & 2 & 48 & inf \\
2 & 0 & 0 & 2 & 1 & 1 & 48 & 270.899 \\
2 & 0 & 0 & 1 & 1 & 1 & 48 & 280.519 \\
2 & 0 & 1 & 1 & 1 & 2 & 48 & inf \\
2 & 0 & 1 & 2 & 1 & 1 & 48 & 273.366 \\
 2 & 0 & 1 & 1 & 1 & 1 & 48 & 276.647 \\
 1 & 0 & 1 & 2 & 1 & 2 & 48 & 237.017 \\
 2 & 0 & 2 & 2 & 1 & 2 & 48 & 237.784 \\
 1 & 0 & 2 & 2 & 1 & 2 & 48 & 236.860 \\
 2 & 0 & 1 & 2 & 1 & 2 & 48 & 233.925 \\ 
\bottomrule
\end{tabularx}
\end{table}

\begin{figure}[H]
	\centering
	\begin{subfigure}{.45\textwidth}
		\includegraphics[width=\textwidth]{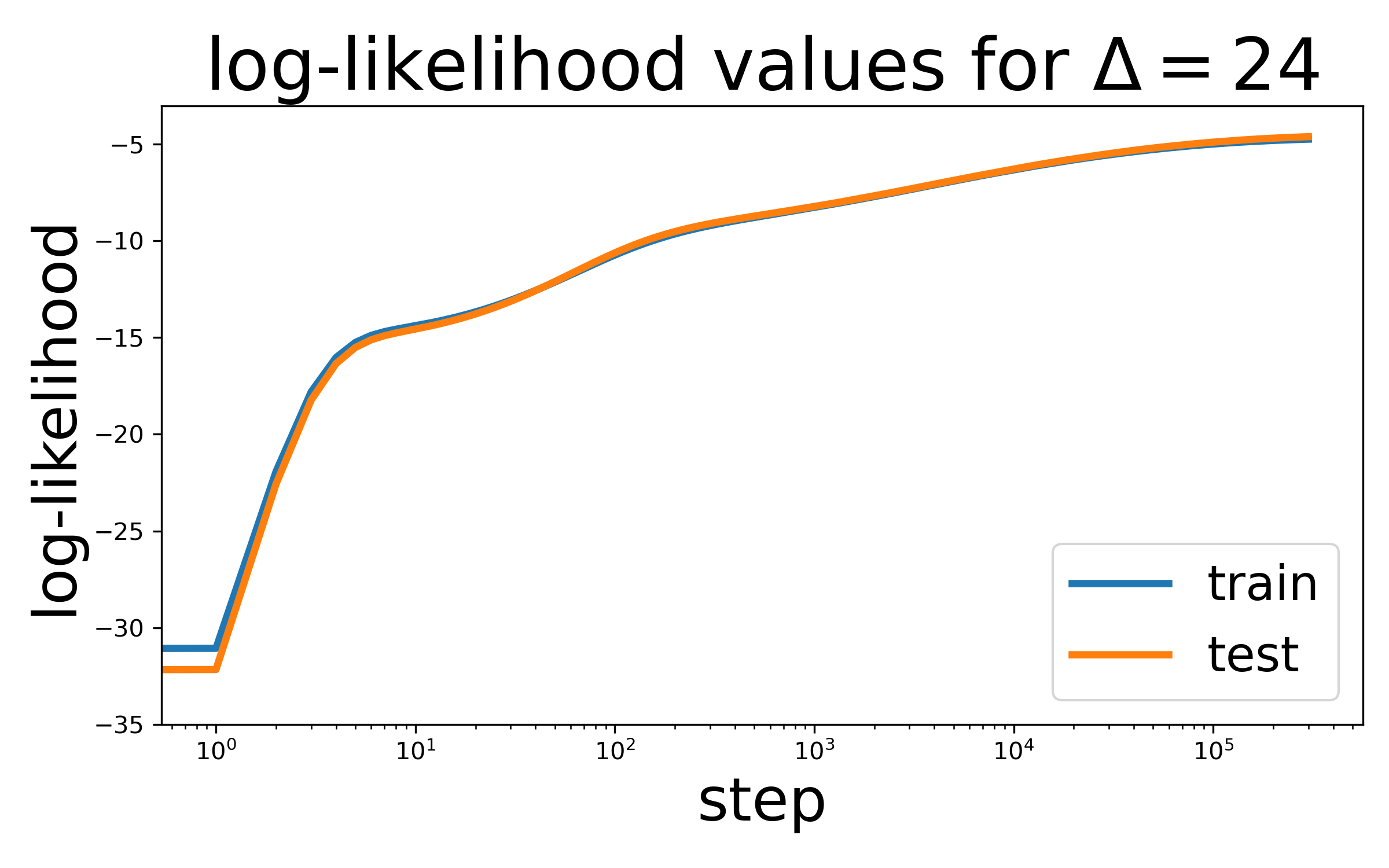}
	\end{subfigure}
	\begin{subfigure}{.45\textwidth}
		\includegraphics[width=\textwidth]{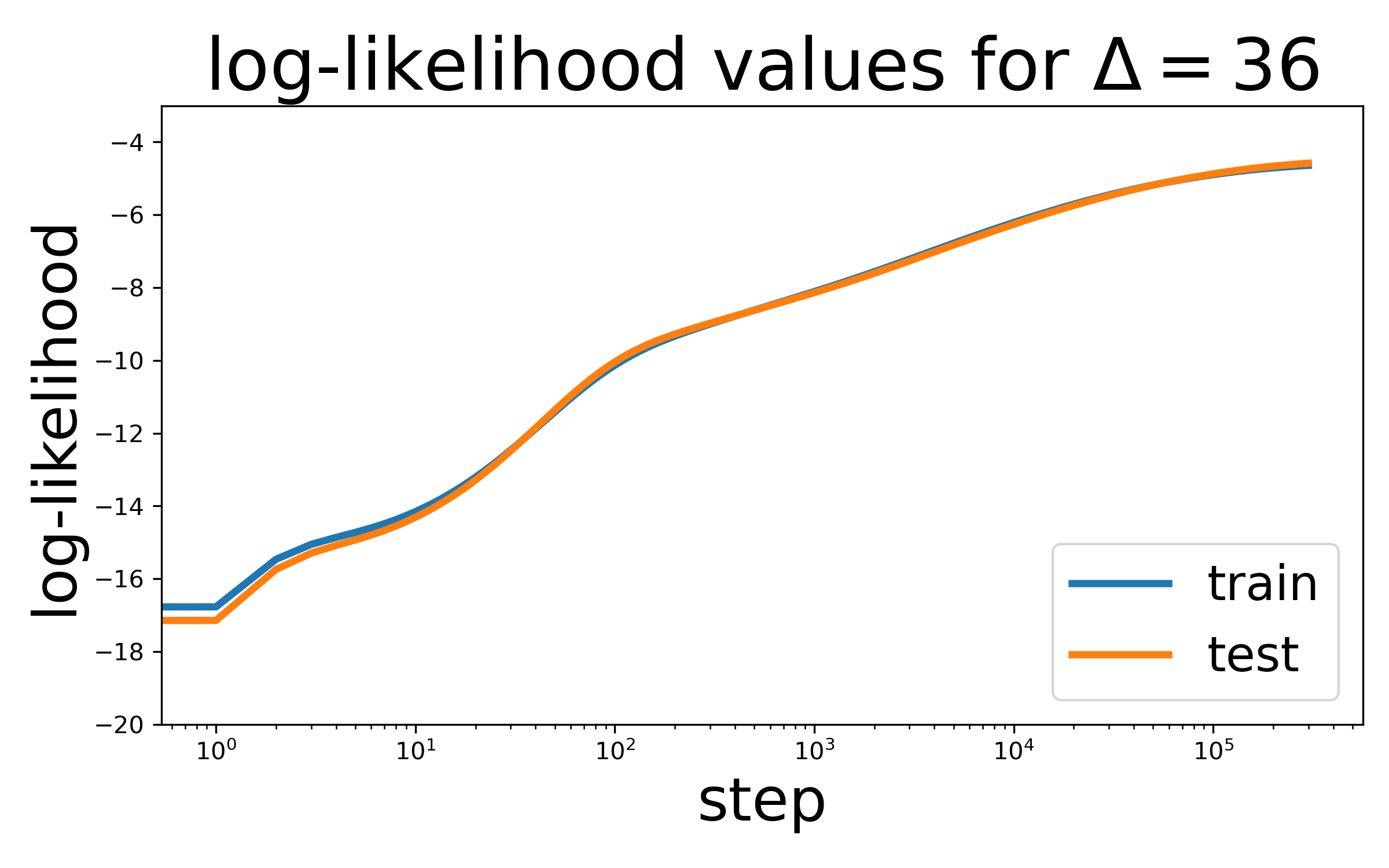}
	\end{subfigure}
	\begin{subfigure}{.45\textwidth}
		\includegraphics[width=\textwidth]{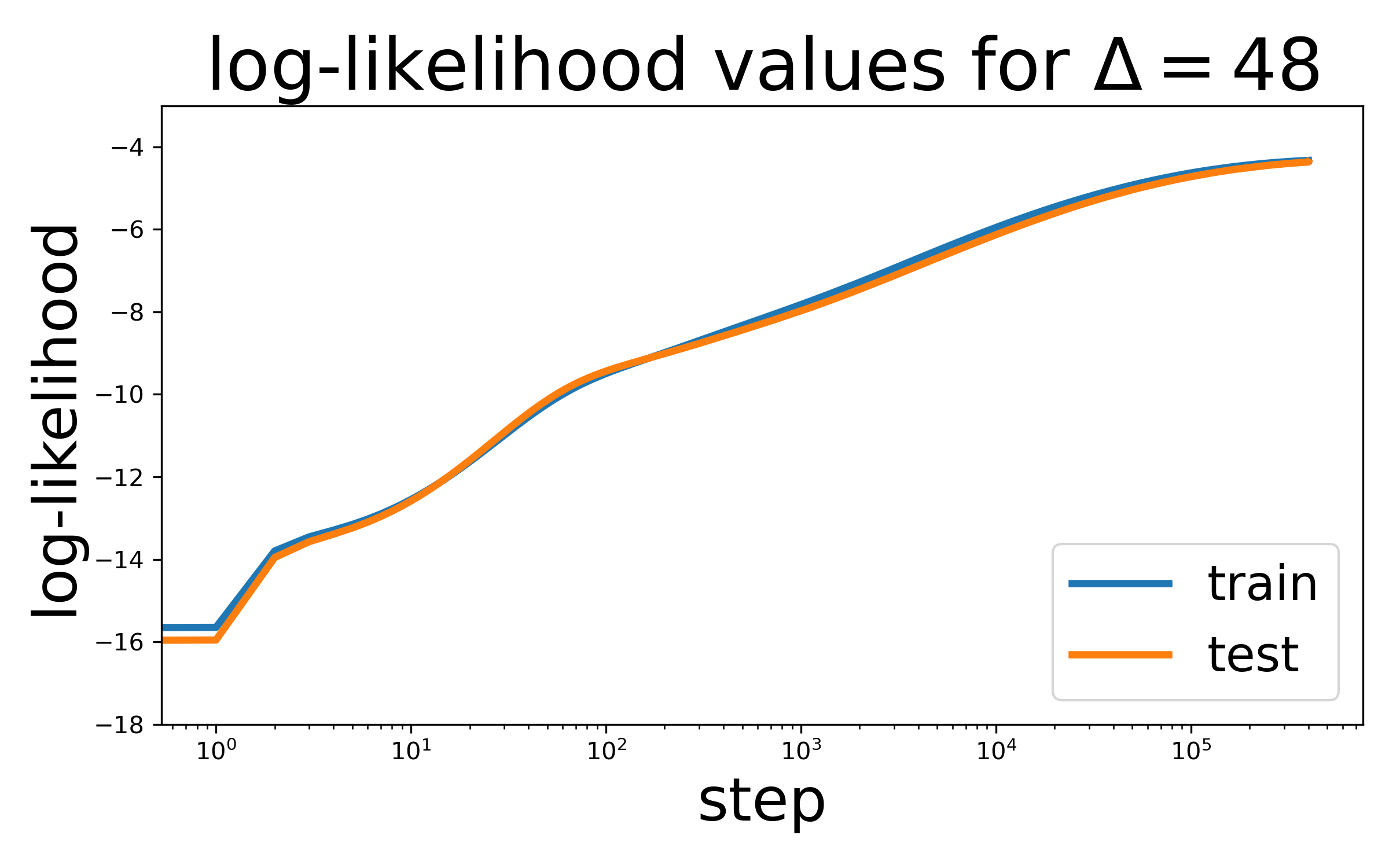}
	\end{subfigure}
	\begin{subfigure}{.45\textwidth}
		\includegraphics[width=\textwidth]{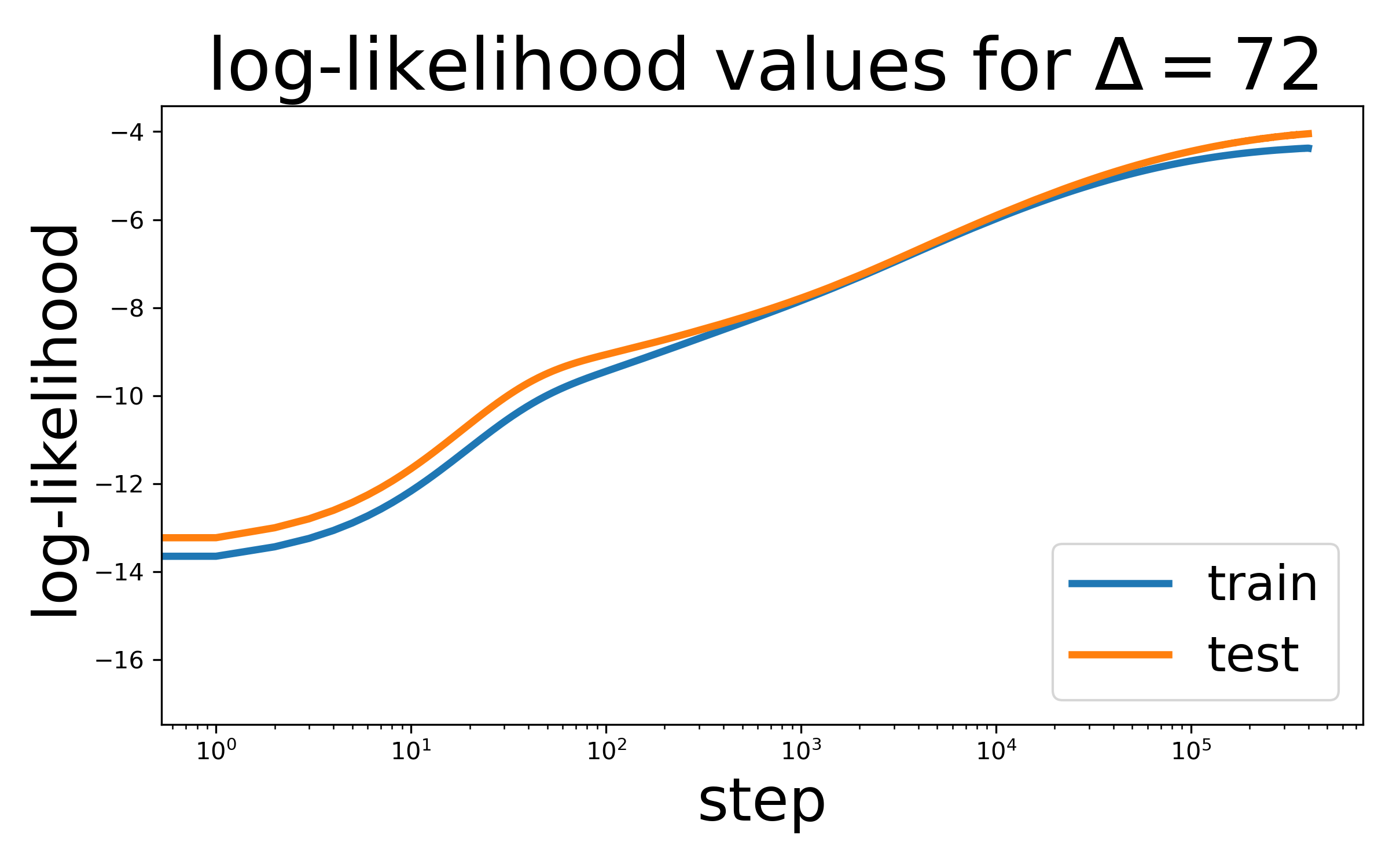}
	\end{subfigure}
    	
	\caption{Log-Likelihood training for different values of $\Delta$.}
	\label{training}
\end{figure}

\begin{figure}[H]
\centering
\includegraphics[width=13cm]{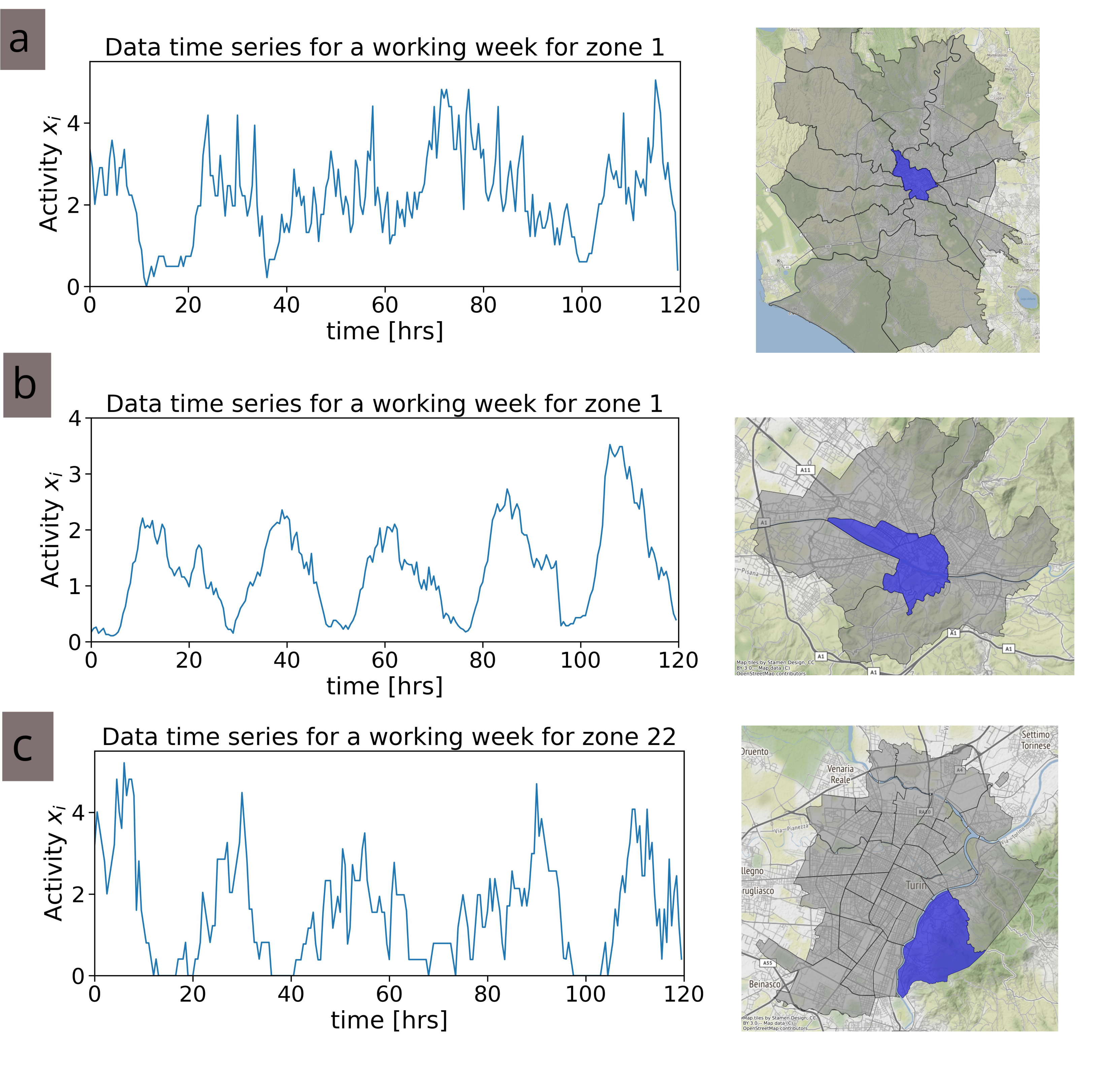}

\caption{Time series activity data for the cities of Rome (a) having $R^2=0.816849$  , Florence (b) having $R^2=0.748687$ and Turin (c) having $0.727507$ .
\label{fig:cities}}
\end{figure}

\bibliography{mybib}  

\end{document}